\documentclass[conference]{IEEEtran}
\IEEEoverridecommandlockouts

\usepackage{cite}
\usepackage{amsmath,amssymb,amsfonts}
\usepackage{algorithmic}
\usepackage{bm}
\usepackage{booktabs}
\usepackage{graphicx}
\usepackage{textcomp}
\usepackage{xcolor}
\usepackage{hyperref}

\newcommand{\emb}[1]{\bm{#1}}
\def\BibTeX{{\rm B\kern-.05em{\sc i\kern-.025em b}\kern-.08em
    T\kern-.1667em\lower.7ex\hbox{E}\kern-.125emX}}

\begin{document}

\title{Hierarchical Multi-Task Learning with Liquidity-Aware Signals for Stock Forecasting}

\author{
\IEEEauthorblockN{
Hengyi Yang\textsuperscript{1,*,\ddag},
Sida Lin\textsuperscript{2,*},
Yiyan Qi\textsuperscript{3,*},
Yankai Chen\textsuperscript{4},\\
Haohan Zhang\textsuperscript{5},
Xianhua Peng\textsuperscript{1,\dag},
Jian Guo\textsuperscript{3,\dag}
}
\IEEEauthorblockA{
\textsuperscript{1}Peking University, Shenzhen, China \quad
\textsuperscript{2}The Chinese University of Hong Kong, Shenzhen, China \\
\textsuperscript{3}International Digital Economy Academy (IDEA), Shenzhen, China \quad
\textsuperscript{4}Cornell University, Ithaca, United States \\
\textsuperscript{5}The Hong Kong University of Science and Technology, Guangzhou, China
}
\IEEEauthorblockA{
hengyiyangyhy@gmail.com, sidalin1@link.cuhk.edu.cn, qiyiyan@idea.edu.cn, yankaichen@acm.org,\\
hzhang760@connect.hkust-gz.edu.cn, xianhuapeng@pku.edu.cn, guojian@idea.edu.cn
}
\IEEEauthorblockA{
\textsuperscript{*}Equal contribution \quad
\textsuperscript{\dag}Corresponding authors \quad
\textsuperscript{\ddag}This work was conducted during his internship at IDEA Research.
}
}

\maketitle

\begin{abstract}
Stock price forecasting is a long-standing challenge in computational finance, driven by the inherent randomness of markets and complex temporal patterns. 
While recent deep-learning models have raised forecasting accuracy by jointly modeling inter-stock and temporal price dynamics, they conflate inter-stock relationships with intra-stock temporal dependencies and focus solely on the univariate objective of price movement. 
To address these limitations, we propose \textbf{LiMT}, a Hierarchical Multi-Task Learning framework that integrates liquidity-aware signals for stock price forecasting. 
LiMT employs a Market Regime Encoder (MRE) module that first extracts contemporaneous cross-stock dependencies, then models each stock's temporal dynamics, yielding a unified latent state.
Building on this latent state, we introduce a Liquidity-Driven Learning (LDL) module, a mixture-of-experts architecture that features cross-task gating mechanisms to jointly predict price movement, volatility, and trading volume. 
We further design an Adaptive Portfolio Optimization (APO) mechanism that converts multi-task forecasts into executable portfolio weights under transaction-cost and liquidity constraints.
Extensive experiments on the CSI300 and CSI500 benchmarks show that LiMT performs best among strong neural and tree-based baselines across the reported metrics. In realistic CSI300 backtests, APO improves annualized return from 3.99\% to 10.01\% and Sharpe ratio from 1.22 to 1.86 over equal weighting, showing that the multi-task forecasts translate into deployable portfolio gains.
\end{abstract}


\begin{center}
\textbf{Code Availability:} An anonymized repository is available at \href{https://anonymous.4open.science/r/LiMT-F039}{anonymous.4open.science/r/LiMT-F039}
\end{center}

\section{Introduction}
Stock price forecasting is a central problem in quantitative finance because it directly affects portfolio construction and risk management. Unlike stationary time series, stock returns are highly non-stationary and jointly influenced by macroeconomic conditions, capital flows, investor sentiment, and unexpected shocks\cite{zhuPredictiveRegressionsMacroeconomic2014,khanMacroeconomicFactorsStock2023}. These dependencies unfold along both temporal and cross-sectional dimensions, making effective forecasting inherently challenging\cite{wang2021hierarchical}.

Learning-based forecasting has advanced from decision trees and ensemble models~\cite{quinlan1986induction,quinlan2014c4,zhou2012ensemble,chen2016xgboost,ke2017lightgbm,zhang2020doubleensemble} to recurrent, convolutional, and Transformer-style architectures that better capture temporal structure in financial signals~\cite{du2021adarnn,wang2022adaptive,zhang2017stock,selvin2017stock,hoseinzade2019cnnpred,vaswani2017attention}. To model inter-stock structure, prior work introduced static relation graphs and later attention-based architectures that dynamically aggregate peer information~\cite{feng2019temporal,wang2021hierarchical,wang2022adaptive,yoo2021accurate,liMASTERMarketGuidedStock2023}. Other extensions incorporate futures or sentiment signals~\cite{xiang2022temporal,linCSPOCrossMarketSynergistic2025}. In parallel, market-microstructure studies show that volume and volatility are important for liquidity, execution, and risk control~\cite{engleMultipleIndicatorsModel2006,nettlesForecastingIntradayVolume2015,szucsForecastingIntradayVolume2017,zhangVolatilityForecastingMachine2022}, yet these signals are still rarely learned jointly with return prediction in end-to-end stock forecasters.

Despite this progress, three structural limitations remain. First, many relation-aware architectures impose a fixed order between temporal aggregation and cross-sectional interaction~\cite{feng2019temporal,wang2021hierarchical,wang2022adaptive,yoo2021accurate,liMASTERMarketGuidedStock2023}. If temporal compression happens first, day-specific peer effects are averaged away before cross-stock interaction; if cross-sectional mixing is followed by temporal aggregation, transient co-movements and stock-specific momentum become entangled. The resulting representation cannot cleanly separate contemporaneous market synchronization from within-stock dynamics. Second, most methods optimize return prediction alone, leaving volume and volatility outside the learning objective even though trading volume carries independent alpha content and volatility directly affects risk-adjusted performance and execution feasibility~\cite{goyenkoTradingVolumeAlpha2024,zhangVolatilityForecastingMachine2022}. Existing multi-task models~\cite{maMultipleStockTime2020a,parkStockMarketForecasting2022} remain largely return-centered and do not explicitly learn liquidity-sensitive auxiliary signals for downstream portfolio construction. Third, portfolio construction is often reduced to simple ranking or threshold rules that ignore liquidity limits, impact, and transaction costs, weakening the link between predictive accuracy and executable returns.

To address these issues, we propose \textbf{LiMT}, a Hierarchical Multi-Task Learning framework that integrates liquidity-aware signals for stock price forecasting. Our contributions are threefold.

\textbf{Market Regime Encoder (MRE)}: We introduce a hierarchical stock transformer that applies cross-stock attention independently at each historical step and only then performs within-stock temporal modeling. This ordering preserves time-specific cross-sectional structure before temporal compression, reducing the signal entanglement induced by rigid sequential pipelines.

\textbf{Liquidity-Driven Learning (LDL)}: We jointly learn return, volume shock, and volatility through a MoE-based multi-task module with cross-task gating, so informative liquidity and risk signals can be transferred to the return objective while conflicting auxiliary information is suppressed.

\textbf{Adaptive Portfolio Optimization (APO)}: We design a lightweight liquidity-aware weighting rule that de-normalizes predicted volume and volatility, linearly fuses liquidity and risk preferences, and improves annualized return from 3.99\% to 10.01\% with Sharpe ratio rising from 1.22 to 1.86 in realistic backtests under participation and transaction-cost constraints.

\section{Problem Formulation}
We study daily cross-sectional stock forecasting with three jointly learned targets aligned with both prediction and execution: next-day return, volume shock, and volatility. On trading day $t$, the model uses the historical feature window to predict the targets realized at $t+2$.

\subsection{Main Task - Return}
The return metric, denoted as ${z}_r$, is computed as the relative change in closing prices over two consecutive time intervals, expressed as ${z}_r = \frac{\mathrm{Close}(t + 2)}{\mathrm{Close}(t + 1)} - 1$. This formula quantifies the percentage variation in closing prices, enabling the simultaneous capture of both the directional movement and magnitude of price changes.

\subsection{Auxiliary Task 1 - Volume Shock}
The volume shock metric, ${z}_v$, measures the deviation of trading volume from its recent historical trend. It is calculated as the difference between the volume at time step $t+2$ and the five-period moving average of volume at time step $t+1$, i.e., ${z}_v = \log\!\left(\mathrm{volume}_{t + 2}\right) 
          - \frac{1}{5}\sum_{i=1}^{5} \log\!\left(\mathrm{volume}_{t+2-i}\right)
$. This metric is designed to identify abnormal trading activities, such as sudden surges in buying or selling pressure.

\subsection{Auxiliary Task 2 - Volatility}
The volatility metric is defined as ${z}_\sigma = \frac{\mathrm{high}_{t + 2} - \mathrm{low}_{t + 2}}{\mathrm{vwap}_{t + 2}}$, which measures price fluctuation intensity. This ratio normalizes the price range (difference between the high and low prices) by the volume-weighted average price (vwap), providing a standardized estimate of the magnitude of price swings within the given time frame.

All targets undergo cross-sectional $Z$-score normalization at each trading day: $ZS(y) = \frac{y - \mu_y}{{std}_y}$, where $\mu_y$ and ${std}_y$ denote the cross-sectional mean and standard deviation of $y$, respectively.

\section{Methodology}
\newcommand{\model}{LiMT}

\begin{figure*}
    \centering
    \includegraphics[width=\textwidth]{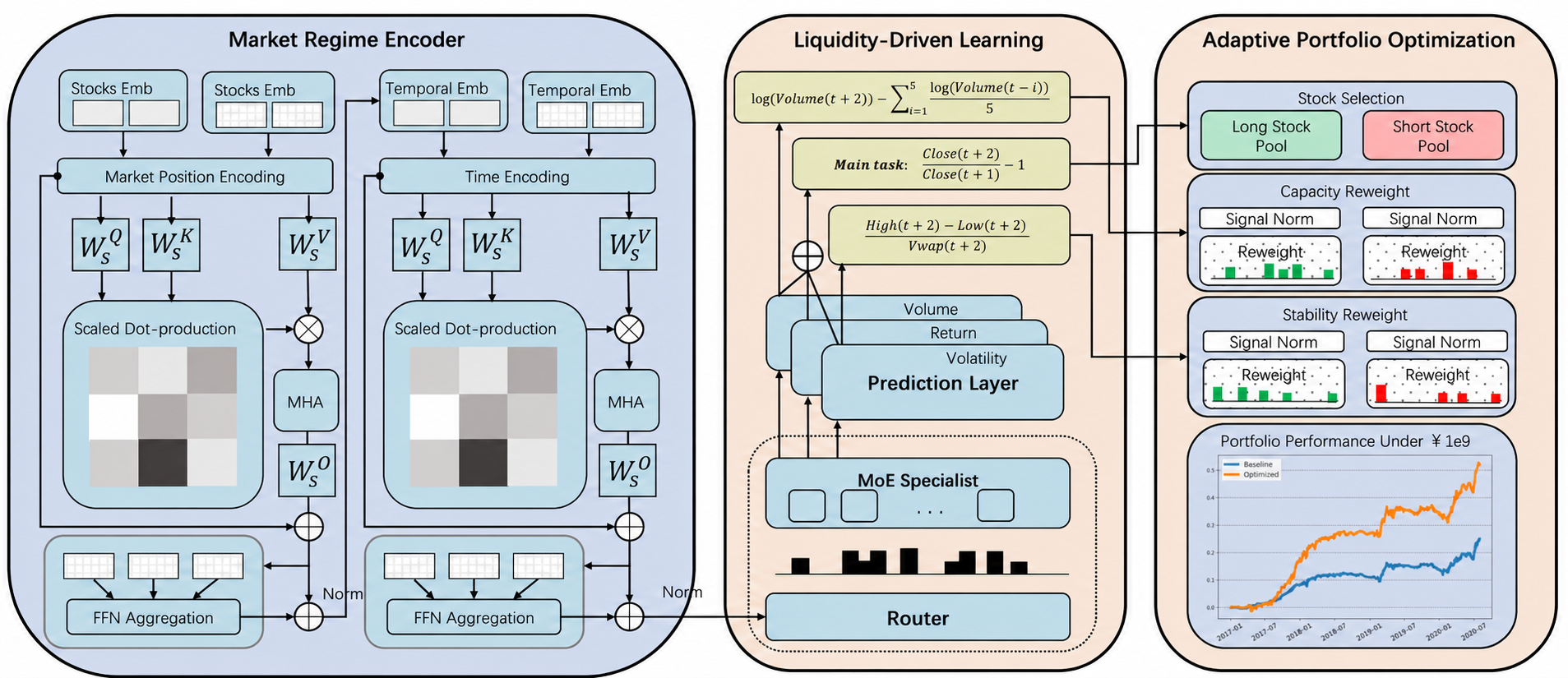}
    \caption{Overview of the \model~framework. Left: MRE module, Middle: LDL module, Right: APO module.}
    \label{fig:framework}
\end{figure*}

\subsection{Overview}
% Capital flows drive price changes, and these flows leave measurable signals in trading volume and volatility. Our goal is to use long-run stock co-movements and persistent temporal links to build a stock price forecasting model that more accurately reflects real price dynamics.
Figure\ref{fig:framework} depicts the architecture of LiMT, which consists of three components. (1) The \textbf{Market Regime Encoder (MRE)} applies cross-stock attention at each historical step and then models within-stock temporal dynamics, so changing market conditions are captured implicitly through time-varying interactions rather than explicit regime labels.
(2) Building on MRE, the \textbf{Liquidity-Driven Learning (LDL)} module jointly predicts return, volume shock, and volatility through shared experts and gated cross-task transfer.
(3) The \textbf{Adaptive Portfolio Optimization (APO)} module then converts these forecasts into time-varying long-short weights under liquidity and participation constraints.

\subsection{Market Regime Encoder Module }
\label{MRE}
To capture both cross-sectional and temporal information, the Market Regime Encoder module comprises two attention-based sub-modules. First, a cross-sectional attention layer enriches each time-step's representation by attending to contemporaneous information across assets. Second, a temporal attention encoder summarizes all information within the rolling window. This orthogonal decomposition enables more precise modeling of both market-wide synchronization and individual stock momentum patterns.

\textit{Feature Projection}: Input features $\emb{X} \in \mathbb{R}^{S \times T \times F}$ are first projected to a $d$-dimensional hidden space to facilitate correlation modeling:
\begin{equation}
\begin{aligned}
\emb{H}^* = \text{Linear}(\emb{X}) \in \mathbb{R}^{S \times T \times d},
\end{aligned}
\end{equation}
where $S$ represents the number of stocks in each trading day, $T$ denotes the rolling window size, and $d$ is the hidden feature dimension.

\subsubsection{\textbf{Cross-Stock Attention}}
To simplify notation, we illustrate the computation for a single time step. For each time step $k$, we extract a cross-sectional slice $\emb{H}_c \in \mathbb{R}^{S \times d}$ from $\emb{H}^*$ to capture contemporaneous correlations across all stocks without accessing future temporal information. The multi-head attention mechanism assigns higher weights to stocks that are more relevant or similar at the given time step:
\begin{equation}
\begin{aligned}
\emb{H}_c = \emb{H}^*[:, k, :] \in \mathbb{R}^{S \times d},
\end{aligned}
\end{equation}
For each attention head $i = 1, ..., h$:
\begin{equation}
\begin{aligned}
\emb{A}_c^{(i)} = \text{Softmax}\left( \frac{(\emb{H}_c \emb{W}^{Q,(i)})(\emb{H}_c \emb{W}^{K,(i)})^\top}{\sqrt{d_k}} \right) \in \mathbb{R}^{S \times S},
\end{aligned}
\end{equation}
\begin{equation}
\begin{aligned}
\emb{H}_{cross}^{(i)} = \emb{A}_c^{(i)} \emb{H}_c \emb{W}^{V,(i)} \in \mathbb{R}^{S \times d_k},
\end{aligned}
\end{equation}
where $\emb{W}^{Q,(i)}, \emb{W}^{K,(i)}, \emb{W}^{V,(i)} \in \mathbb{R}^{d \times d_k}$ are projection matrices for head $i$ with $d_k = d/h$ and $\emb{H}_{cross}^{(i)}$ is the hidden states of heads. 

Multi-head concatenation and output projection given as:
\begin{equation}
\begin{aligned}
\emb{H}^{k}_{cross} = \text{Concat}(\emb{H}_{cross}^{(1)}, ..., \emb{H}_{cross}^{(h)}) \emb{W}^O \in \mathbb{R}^{S \times d},
\end{aligned}
\end{equation}
where $\emb{W}^O \in \mathbb{R}^{d \times d}$ is the output projection matrix and $\emb{H}^{k}_{cross}$ denotes the hidden state at the ${k}$-th time step within the look-back window, with k $\in \{1,2,...,T\}$. 

\textit{Residual Connection and Normalization}: To stabilize training and preserve information flow, we apply residual connections with layer normalization:
\begin{equation}
\begin{aligned}
\emb{H}^{k}_{cross} = \text{LayerNorm}(\emb{H}_c + \emb{H}^{k}_{cross}) \in \mathbb{R}^{S \times d}.
\end{aligned}
\end{equation}
Hidden state with cross-stock information given as:
\begin{equation}
\begin{aligned}
\emb{H}^* = \text{Concat}(\emb{H}^{k}_{cross}) \in \mathbb{R}^{S \times T \times d}
\end{aligned}
\end{equation}
% where $\emb{H}^{k}_{cross}$ denotes the hidden state at the ${k}$-th time step within the look-back window, with k $\in \{1,2,...,T\}$ and $\emb{H}^*$ is the concatenation of all time steps.

\subsubsection{\textbf{Temporal-Stock Modeling}}
After capturing cross-stock dependencies, we model temporal dynamics within each stock to extract time-series patterns and long-range dependencies. To simplify notation, we illustrate the computation for a single $q$-th stock. 
For each stock $q$, we extract a temporal slice $\emb{H}_t \in \mathbb{R}^{T \times d}$ from the cross-stock enhanced features to capture dependencies between past and recent time steps:
\begin{equation}
\begin{aligned}
\emb{H}_t = \emb{H}^*[q, :, :] \in \mathbb{R}^{T \times d},
\end{aligned}
\end{equation}
For each attention head $i = 1, ..., h$:
\begin{equation}
\begin{aligned}
\emb{A}_t^{(i)} = \text{Softmax}\left( \frac{(\emb{H}_t \tilde{\emb{W}}^{Q,(i)})(\emb{H}_t \tilde{\emb{W}}^{K,(i)})^\top}{\sqrt{d_k}} \right) \in \mathbb{R}^{T \times T},
\end{aligned}
\end{equation}
\begin{equation}
\begin{aligned}
\emb{H}_{t}^{(i)} = \emb{A}_t^{(i)} \emb{H}_t \tilde{\emb{W}}^{V,(i)} \in \mathbb{R}^{T \times d_k},
\end{aligned}
\end{equation}
where $\tilde{\emb{W}}^{Q,(i)}, \tilde{\emb{W}}^{K,(i)}, \tilde{\emb{W}}^{V,(i)} \in \mathbb{R}^{d \times d_k}$ are temporal projection matrices for head $i$ and  $\emb{H}_{t}^{(i)}$ is the output of different heads. 

Multi-head concatenation and output projection given as:
\begin{equation}
\begin{aligned}
\emb{H}^{q}_{temporal} = \text{Concat}(\emb{H}_{t}^{(1)}, ..., \emb{H}_{t}^{(h)}) \tilde{\emb{W}}^O \in \mathbb{R}^{T \times d},
\end{aligned}
\end{equation}
where $\tilde{\emb{W}}^O \in \mathbb{R}^{d \times d}$ is the temporal output projection matrix and $\emb{H}^{q}_{temporal}$ denotes the hidden state of ${q}$-th stock of $T$s time step, with $q$ $\in \{1,2,...,S\}$.

\textit{Residual Connection and Normalization}: Similar to cross-stock modeling, we apply residual connections:
\begin{equation}
\begin{aligned}
\emb{H}^{q}_{temporal} = \text{LayerNorm}(\emb{H}_t + \emb{H}^{q}_{temporal}).
\end{aligned}
\end{equation}
The representation for stock $q$ uses the most recent temporal information:
\begin{equation}
\begin{aligned}
h^{q} = \emb{H}^{q}_{temporal}[-1, :] \in \mathbb{R}^d,
\end{aligned}
\end{equation}
For the representation of all the stocks at ${T}$-th time given as:
\begin{equation}
\begin{aligned}
\emb{h}^* = \text{Concat}(h^{q}) \in \mathbb{R}^{S \times d}.
\end{aligned}
\end{equation}
where $h^{q}_s$ denotes the hidden state of ${q}$-th stock at the last time step, with $q$ $\in \{1,2,...,S\}$ and $\emb{h}^*$ is the concatenation of all stocks.

\subsection{Liquidity-Driven Learning (LDL) Module}
\label{LDL}
Capital flows are primarily reflected in changes in liquidity, including price, trading volume, and volatility. These three indicators capture complementary aspects of market dynamics and often exhibit regime-dependent behaviors. To jointly model these signals under varying market conditions, we introduce the Liquidity-Driven Learning (LDL) module. It integrates a Multi-gate Mixture-of-Experts architecture with a cross-task knowledge transfer mechanism to enhance return forecasting with liquidity-aware supervision.

\subsubsection{\textbf{Multi-gate Mixture-of-Experts (MMoE)}}
The MMoE module consists of expert routing and task-specific expert aggregation across multiple prediction objectives.

\textit{Expert Routing}: Each task $k \in \{r, v, \sigma\}$, corresponding to return forecasting, volume anomaly prediction, and volatility estimation, is governed by an independent gating function:
\begin{equation}
\begin{aligned}
    \alpha^k = \text{Softmax}(\Phi_k(\emb{h}^*)) \in \mathbb{R}^{S \times n},
\end{aligned}
\end{equation}
where $\Phi_k$ denotes a task-conditioned routing function that learns the most relevant combinations of experts for each objective, and $n$ is the number of experts.

\textit{Expert Aggregation}: The task-specific representation is computed as a weighted combination of expert outputs:
\begin{equation}
\begin{aligned}
    o_k = \sum_{e=1}^{n} \alpha^k[e] \cdot \text{Expert}_e(\emb{h}^*) \in \mathbb{R}^{S \times d}.
\end{aligned}
\end{equation}

\textit{Feature Transformation}: Before cross-task transfer, each task representation is projected into a shared latent space through dimensionality reduction and non-linear activation:
\begin{equation}
\begin{aligned}
G_k(o_k) = \text{ReLU}(\text{Linear}_k(o_k)) \in \mathbb{R}^{S \times \frac{d}{2}},
\end{aligned}
\end{equation}
where $k \in \{r, v, \sigma\}$ and $\text{Linear}_k: \mathbb{R}^{S \times d} \rightarrow \mathbb{R}^{S \times \frac{d}{2}}$ denotes a task-specific projection layer.

\subsubsection{\textbf{Cross-Task Knowledge Transfer}}
To leverage auxiliary liquidity signals for improving return prediction, we introduce a stock-wise gating mechanism that adaptively transfers information from volume and volatility tasks to the primary return objective. This design reflects the financial intuition that volume shocks and volatility patterns may provide leading indicators of future price movements, while their usefulness varies across assets and market regimes.

The transfer strength from each auxiliary task is controlled by an independent scalar gate:
\begin{equation}
\begin{aligned}
\beta_v &= Sigmoid\bigl(\text{Linear}_{v \rightarrow r}(G_v(o_v))\bigr) \in \mathbb{R}^{S \times 1}, \\
\beta_\sigma &= Sigmoid\bigl(\text{Linear}_{\sigma \rightarrow r}(G_\sigma(o_\sigma))\bigr) \in \mathbb{R}^{S \times 1},
\end{aligned}
\end{equation}
where $\text{Linear}_{v \rightarrow r}(\cdot)$ and $\text{Linear}_{\sigma \rightarrow r}(\cdot)$ are two distinct projection functions. The coefficient $\beta_v$ determines when volume-related liquidity signals should be emphasized for return forecasting, whereas $\beta_\sigma$ controls the contribution of volatility-driven risk information. By separating these two gates, the model can capture heterogeneous cross-task relationships and selectively inject auxiliary knowledge when it is predictive, thereby reducing negative transfer.

\textit{Enhanced Representation}: The return representation is enhanced through gated fusion of auxiliary task features:
\begin{equation}
\begin{aligned}
o'_r = G_r(o_r) + \beta_v \odot G_v(o_v) + \beta_\sigma \odot G_\sigma(o_\sigma) \in \mathbb{R}^{S \times \frac{d}{2}},
\end{aligned}
\end{equation}
where $\odot$ denotes stock-wise broadcasting along the feature dimension. This formulation enables adaptive incorporation of complementary liquidity information into the return forecasting process.

\textit{Prediction}: Each task employs a dedicated output layer to generate the final forecasts:
\begin{equation}
\begin{aligned}
    \hat{z}_r &= \text{Linear}_r(o'_r) \in \mathbb{R}^{S \times 1}, \\
    \hat{z}_v &= \text{Linear}_v(G_v(o_v)) \in \mathbb{R}^{S \times 1}, \\
    \hat{z}_\sigma &= \text{Linear}_\sigma(G_\sigma(o_\sigma)) \in \mathbb{R}^{S \times 1}.
\end{aligned}
\end{equation}
where $\hat{z}_r$, $\hat{z}_v$, and $\hat{z}_\sigma$ denote the predicted return, volume anomaly, and volatility values, respectively. The return prediction is computed from the enhanced representation $o'_r$ with cross-task knowledge transfer, while auxiliary tasks are predicted from their own transformed representations.

\subsubsection{\textbf{Multi-task Loss}} We optimize a weighted multi-task loss function using Mean Squared Error (MSE) on z-score normalized targets:
\begin{equation}
\begin{aligned}
    \mathcal{L} = \sum_{i \in \{r,v,\sigma\}} \lambda_i \cdot \text{MSE}(\hat{z}_i, z_i).
\end{aligned}
\end{equation}
where $\lambda_r, \lambda_v, \lambda_\sigma$ are task-specific weights that balance the contribution of each loss item during training.

\subsection{Adaptive Portfolio Optimization Module}
\label{APO}
Building on LDL, APO converts multi-task forecasts into executable long-short weights. At trading day $t$, stocks are sorted by $\hat{z}_r$, and the top and bottom deciles define the long and short candidate sets. For a candidate set $\mathcal{G}_t$, the auxiliary outputs are first de-normalized to physical units:
\begin{equation}
\begin{aligned}
    \hat{A}_{i,t+2} &= \exp\!\left(\hat{z}_{v,i,t} \cdot s^V_t + \mu^V_t + \bar{\ell}_{i,t}\right) \cdot p^{\text{vwap}}_{i,t}, \\
    \hat{\sigma}_{i,t+2} &= \hat{z}_{\sigma,i,t} \cdot s^\sigma_t + \mu^\sigma_t,
\end{aligned}
\end{equation}
where $\mu^V_t, s^V_t$ and $\mu^\sigma_t, s^\sigma_t$ are the day-$t$ cross-sectional moments of volume shock and volatility, $\bar{\ell}_{i,t}=\frac{1}{5}\sum_{k=0}^{4}\log V_{i,t-k}$ is the 5-day mean of log-volume, and $p^{\text{vwap}}_{i,t}$ is the day-$t$ VWAP.

We then apply the linear normalization
\begin{equation}
\begin{aligned}
    D_{\mathcal{G}_t}(x) = \sum_{k\in\mathcal{G}_t}(x_k-\min_{j\in\mathcal{G}_t} x_j),
\end{aligned}
\end{equation}
\begin{equation}
\mathrm{Norm}_{\mathcal{G}_t}(x_i)=
\begin{cases}
\dfrac{x_i-\min_{j\in\mathcal{G}_t} x_j}{D_{\mathcal{G}_t}(x)}, & \text{if } D_{\mathcal{G}_t}(x)>0,\\[8pt]
\dfrac{1}{|\mathcal{G}_t|}, & \text{otherwise},
\end{cases}
\end{equation}
to form liquidity and risk weights:
\begin{equation}
\begin{aligned}
    w_i^A &= \mathrm{Norm}_{\mathcal{G}_t}\!\left(\hat{A}_{i,t+2}\right), \\
    w_i^\sigma &= \mathrm{Norm}_{\mathcal{G}_t}\!\left(-\hat{\sigma}_{i,t+2}\right).
\end{aligned}
\end{equation}
The final executable weight is
\begin{equation}
\begin{aligned}
    w_i^* = \min\!\left(r \cdot w_i^\sigma + (1-r) \cdot w_i^A,\ \frac{\alpha \cdot A_{i,t}}{C}\right),
\end{aligned}
\end{equation}
where $r \in [0,1]$ balances risk control and liquidity preference, $\alpha=2\%$ is the participation-rate cap, $A_{i,t}$ is realized turnover, and $C$ is the portfolio capital.

\section{Experiments}

\begin{table*}[t]
\centering
\renewcommand\arraystretch{1.1}
\footnotesize
\caption{Performance comparison on CSI300 and CSI500 data. Bold = best, underline = second-best.}
\label{tab:public_performance}
\tabcolsep=0.05cm
\begin{tabular}{l|cccc|cccc}
\toprule
\textbf{Model} &
\multicolumn{4}{c|}{\textbf{CSI300}} &
\multicolumn{4}{c}{\textbf{CSI500}} \\
\midrule
& \textbf{IC}$\uparrow$ & \textbf{ICIR}$\uparrow$ & \textbf{AR$_{\text{excess}}$}$\uparrow$ & \textbf{IR$_{\text{excess}}$}$\uparrow$ &
  \textbf{IC}$\uparrow$ & \textbf{ICIR}$\uparrow$ & \textbf{AR$_{\text{excess}}$}$\uparrow$ & \textbf{IR$_{\text{excess}}$}$\uparrow$ \\
\midrule
LSTM & 0.0321$\pm$0.01 & 0.2267$\pm$0.04 & 0.0381$\pm$0.03 & 0.5561$\pm$0.46 & 0.0381$\pm$0.00 & 0.3636$\pm$0.02 & 0.0596$\pm$0.04 & 1.0243$\pm$0.64 \\
GRU & 0.0324$\pm$0.00 & 0.2448$\pm$0.05 & 0.0344$\pm$0.02 & 0.5160$\pm$0.25 & 0.0413$\pm$0.00 & \underline{0.3962}$\pm$0.03 & 0.0723$\pm$0.01 & 1.1982$\pm$0.12 \\
Transformer & 0.0243$\pm$0.00 & 0.1879$\pm$0.01 & 0.0286$\pm$0.00 & 0.4317$\pm$0.10 & 0.0258$\pm$0.00 & 0.2296$\pm$0.03 & 0.0392$\pm$0.02 & 0.6317$\pm$0.26 \\
ALSTM & 0.0369$\pm$0.00 & 0.2756$\pm$0.03 & 0.0645$\pm$0.04 & 0.9128$\pm$0.55 & 0.0406$\pm$0.00 & 0.3741$\pm$0.04 & 0.0473$\pm$0.02 & 0.7839$\pm$0.26 \\
SFM & 0.0370$\pm$0.00 & 0.2879$\pm$0.04 & 0.0465$\pm$0.02 & 0.5672$\pm$0.29 & 0.0353$\pm$0.00 & 0.3007$\pm$0.04 & 0.0310$\pm$0.00 & 0.2728$\pm$0.03 \\
TCN & 0.0279$\pm$0.00 & 0.2181$\pm$0.01 & 0.0262$\pm$0.02 & 0.4133$\pm$0.25 & 0.0050$\pm$0.00 & 0.0587$\pm$0.00 & -0.0682$\pm$0.02 & -1.0386$\pm$0.26 \\
TabNet & 0.0283$\pm$0.01 & 0.2168$\pm$0.06 & 0.0278$\pm$0.03 & 0.3767$\pm$0.46 & 0.0341$\pm$0.01 & 0.3187$\pm$0.08 & -0.0016$\pm$0.06 & 0.0477$\pm$0.97 \\
TFT & 0.0358$\pm$0.00 & 0.2169$\pm$0.03 & 0.0847$\pm$0.02 & 0.8131$\pm$0.19 & 0.0398$\pm$0.01 & 0.2900$\pm$0.04 & 0.0117$\pm$0.01 & 0.0894$\pm$0.09 \\
Localformer & 0.0321$\pm$0.01 & 0.2469$\pm$0.05 & 0.0225$\pm$0.03 & 0.3310$\pm$0.50 & 0.0313$\pm$0.00 & 0.2827$\pm$0.02 & 0.0250$\pm$0.02 & 0.4120$\pm$0.32 \\
MASTER & 0.0496$\pm$0.01 & 0.4321$\pm$0.00 & 0.1035$\pm$0.01 & 1.2104$\pm$0.01 & \underline{0.0435}$\pm$0.00 & 0.3862$\pm$0.03 & \underline{0.1242}$\pm$0.00 & 1.3415$\pm$0.00 \\
\midrule
XGBoost & \underline{0.0538}$\pm$0.00 & 0.3979$\pm$0.00 & 0.0820$\pm$0.00 & 0.9170$\pm$0.00 & 0.0410$\pm$0.00 & 0.3390$\pm$0.00 & 0.0603$\pm$0.00 & 0.7475$\pm$0.00 \\
CatBoost & 0.0494$\pm$0.00 & 0.3467$\pm$0.01 & 0.0473$\pm$0.00 & 0.3507$\pm$0.01 & 0.0423$\pm$0.00 & 0.3349$\pm$0.00 & 0.0604$\pm$0.00 & 0.7603$\pm$0.00 \\
LightGBM & 0.0455$\pm$0.00 & 0.3674$\pm$0.00 & 0.1043$\pm$0.00 & 1.2268$\pm$0.00 & 0.0348$\pm$0.00 & 0.3387$\pm$0.00 & 0.1186$\pm$0.00 & \underline{1.5247}$\pm$0.00 \\
DoubleEnsemble & 0.0521$\pm$0.00 & \underline{0.4223}$\pm$0.01 & \underline{0.1158}$\pm$0.01 & \underline{1.3432}$\pm$0.11 & 0.0408$\pm$0.00 & 0.3710$\pm$0.01 & 0.0382$\pm$0.00 & 0.1723$\pm$0.00 \\
\midrule
\textbf{\model~} & \textbf{0.0575}$\pm$0.00 & \textbf{0.4462}$\pm$0.03 & \textbf{0.1809}$\pm$0.00 & \textbf{2.0414}$\pm$0.03 & \textbf{0.0496}$\pm$0.00 & \textbf{0.452}$\pm$0.02 & \textbf{0.1529}$\pm$0.00 & \textbf{2.0166}$\pm$0.03 \\
\textbf{Relative Improvement (\%)} & \textbf{6.88\%} & \textbf{5.41\%} & \textbf{56.22\%} & \textbf{51.98\%} & \textbf{13.94\%} & \textbf{14.08\%} & \textbf{23.11\%} & \textbf{32.26\%} \\
\bottomrule
\end{tabular}
\end{table*}

\begin{table*}[t]
\centering
\renewcommand\arraystretch{1.1}
\small
\caption{Ablation Study on CSI300 data.}
\label{tab:ablation}
\begin{tabular}{l|cccc}
\toprule
\textbf{Model} & \textbf{IC}$\uparrow$ & \textbf{ICIR}$\uparrow$ & \textbf{AR$_{\text{excess}}$}$\uparrow$ & \textbf{IR$_{\text{excess}}$}$\uparrow$ \\
\midrule
w/o Auxiliary Tasks & $0.0543 \pm 0.00$ & $0.4035 \pm 0.00$ & $0.1734 \pm 0.00$ & $1.9418 \pm 0.01$ \\
w/o LDL & $0.0568 \pm 0.00$ & $0.4342 \pm 0.01$ & $0.1845 \pm 0.01$ & $1.9993 \pm 0.00$ \\
w/o Auxiliary Tasks \& LDL & $0.0528 \pm 0.00$ & $0.3841 \pm 0.02$ & $0.1687 \pm 0.00$ & $1.8104 \pm 0.00$ \\
w/o Cross-Stock Modeling & $0.0493 \pm 0.00$ & $0.4130 \pm 0.00$ & $0.1277 \pm 0.01$ & $1.6929 \pm 0.01$ \\
w/o Auxiliary Tasks \& LDL \& Cross-Stock Modeling & $0.0489 \pm 0.00$ & $0.4060 \pm 0.00$ & $0.1049 \pm 0.03$ & $1.3838 \pm 0.02$ \\
\bottomrule
\end{tabular}
\end{table*}

In this section, we conduct experiments to answer the following three research questions:
\begin{itemize}
    \item \textbf{RQ1} How does \model~compare with strong neural and tree-based baselines on real-world stock forecasting benchmarks?
    \item \textbf{RQ2} Which components of the proposed architecture drive the predictive gains?
    \item \textbf{RQ3} Do the auxiliary tasks and APO improve executable portfolio performance under realistic trading constraints?
    % \item \textbf{RQ4} Does our model still work if we transfer it to different trading frequency dataset?
\end{itemize}

\subsubsection{\textbf{Datasets}}
We evaluate \model~on the Chinese A-share market using CSI300 and CSI500 constituents. CSI300 covers the largest and most liquid stocks, while CSI500 extends evaluation to a broader mid-cap universe. The data span 2008--2020, with training/validation/test splits of 2008-01-01 to 2014-12-31, 2015-01-01 to 2016-12-31, and 2017-01-01 to 2020-08-01. We use the public Alpha158 feature set from Qlib~\cite{yang2020qlib}, with lookback window $\tau=21$ and prediction horizon $w=1$ day.

\subsubsection{\textbf{Baselines}}
We compare \model~against fourteen representative baselines. The deep-learning group includes LSTM~\cite{lstm}, GRU~\cite{gru}, Transformer~\cite{vaswani2017attention}, ALSTM~\cite{qin_dual-stage_2017}, SFM~\cite{sfm}, TCN~\cite{tcn}, TabNet~\cite{tabnet}, TFT~\cite{TFT}, Localformer~\cite{yadati2025localformer}, and MASTER~\cite{liMASTERMarketGuidedStock2023}. The tree-based group includes XGBoost~\cite{chen2016xgboost}, CatBoost~\cite{catboost}, LightGBM~\cite{ke2017lightgbm}, and DoubleEnsemble~\cite{zhang2020doubleensemble}.

\subsubsection{\textbf{Evaluation}}
We evaluate both predictive quality and trading value using IC, ICIR, excess annualized return, and excess information ratio; precise metric definitions are summarized in the appendix. The portfolio simulation follows a daily top-50 long strategy with 20 replacements per day.

\subsubsection{\textbf{Implementation}}
We implement \model~using PyTorch on the Qlib platform~\cite{yang2020qlib}. Key hyperparameters are $d=256$, $lr=10^{-5}$, $head_{cross}=4$, and $head_{temp}=2$. Models are trained for up to 100 epochs with early stopping. Each experiment is repeated three times and reported as mean $\pm$ standard deviation. Complete hyperparameter tuning details, baseline configurations, and backtest setup are provided in Appendix~\ref{sec:EI}.

\subsection{Overall Performance (RQ1)}
Table~\ref{tab:public_performance} presents the comprehensive performance comparison across all baseline methods on both CSI300 and CSI500 datasets.

\model~ranks first on all four metrics across both datasets. The gains are especially pronounced on the portfolio-oriented metrics: on CSI300, excess annualized return and excess information ratio improve by 56.22\% and 51.98\% over the strongest baseline, while on CSI500 the corresponding gains are 23.11\% and 32.26\%. Tree-based models, especially XGBoost, remain competitive on IC and ICIR, but \model~translates predictions into stronger excess returns, suggesting that jointly modeling cross-sectional structure and liquidity-aware auxiliary signals is particularly useful for deployable decisions. Performance is also slightly more stable on CSI300 than on CSI500, which is consistent with the higher liquidity and lower heterogeneity of large-cap constituents.

\textbf{Interpretability Analysis of Attention and Routing Mechanisms:}
To illustrate how the framework uses cross-sectional, temporal, and task-specific structure, we conduct a case study on March 20, 2020 for the target stock \texttt{SH600000}. We visualize the cross-stock attention patterns produced by MRE, the cross-time propagation scores induced by temporal and cross-sectional attention, and the task-specific routing distributions over the shared LDL expert pool. For readability, we average attention weights across heads to obtain a cross-stock matrix $\emb{A}_{c,j}$ at each historical step $j$ and a temporal matrix $\emb{A}_{t,u}$ for each stock $u$.

\begin{figure}[t]
\centering
\includegraphics[width=1.0\columnwidth]{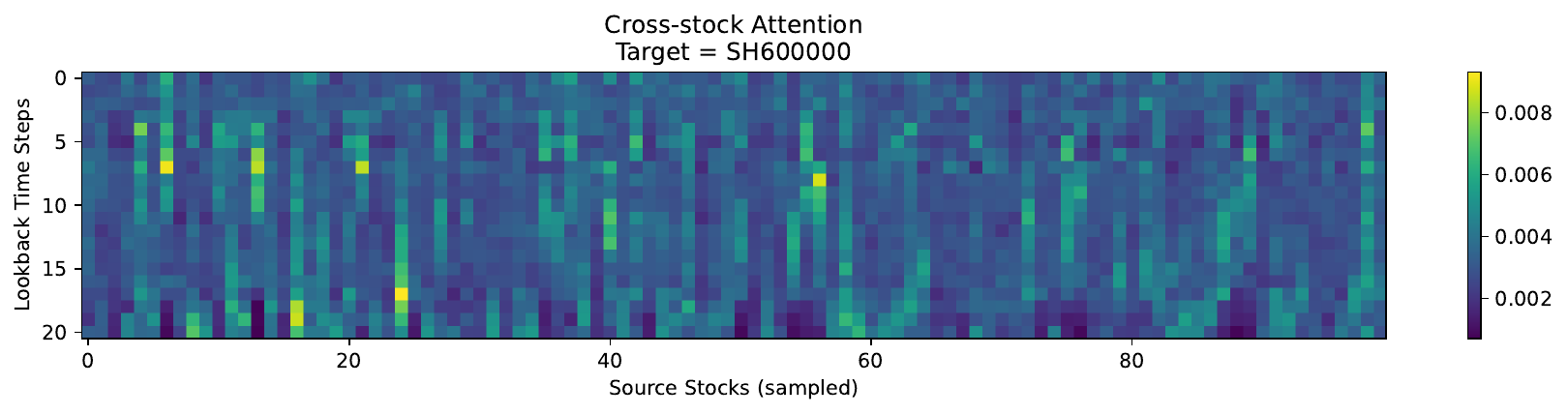} 
\caption{Cross-stock attention heatmap. The x-axis shows 100 sampled stocks, and the y-axis shows the 21-day lookback window on March 20, 2020.}
\label{fig:cross_stock_strip}
\end{figure}

Figure~\ref{fig:cross_stock_strip} shows that the cross-stock attention of \texttt{SH600000} over 100 randomly sampled peers is sparse and selective: only a small subset of stocks receives consistently high weights, and the attended peers vary across historical steps. This pattern is consistent with the view that relevant cross-sectional dependencies are time-varying rather than static.

\begin{figure}[t]
\centering
\includegraphics[width=1.0\columnwidth]{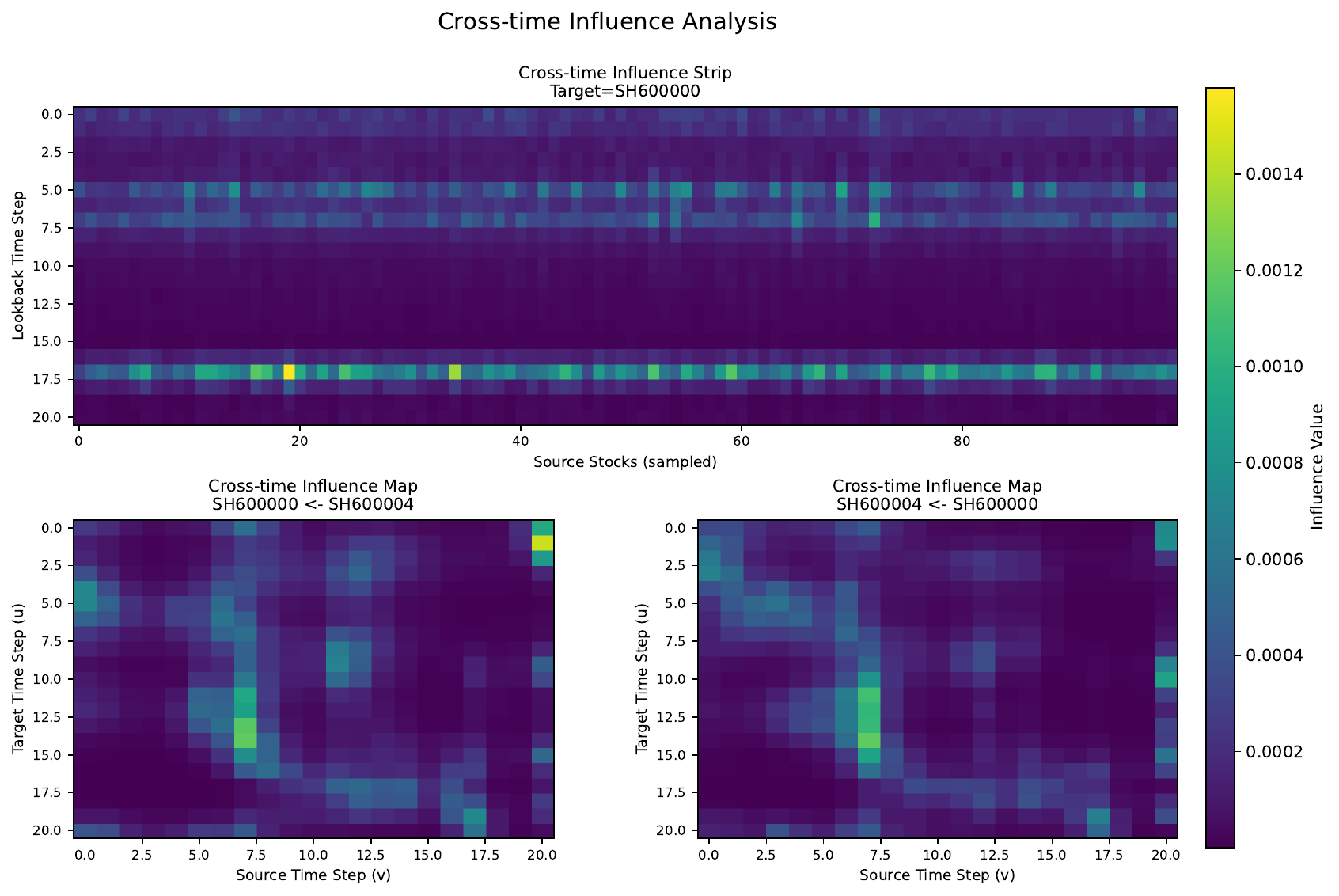} 
\caption{Cross-time propagation heatmap at the last timestamp. The x-axis shows 100 sampled stocks, and the y-axis shows the 21-day lookback window on March 20, 2020.}
\label{fig:cross_time_strip}
\end{figure}

Figure~\ref{fig:cross_time_strip} further examines how temporal importance interacts with cross-sectional relations through cross-time propagation scores. Concretely, for a target stock $u$ and a source stock $v$, we define the propagation strength at historical step $j$ as
\begin{equation}
M[j,v] = \emb{A}_{t,u}[-1,j]\cdot \emb{A}_{c,j}[u,v],
\end{equation}
where $\emb{A}_{t,u}[-1,j]$ denotes the temporal attention weight assigned by stock $u$ when forecasting the last step of the rolling window, and $\emb{A}_{c,j}[u,v]$ represents the cross-stock attention weight from stock $v$ to stock $u$ at the $j$-th historical step. The first row of Figure~\ref{fig:cross_time_strip} visualizes $M[j,v]$ over the lookback window and shows that peer influence is concentrated on a few historical steps rather than being spread uniformly across time.

The second row of Figure~\ref{fig:cross_time_strip} reports cross-time propagation maps between \texttt{SH600000} and \texttt{SH600004}, alternately treating each stock as source and target. The patterns are asymmetric and not strictly diagonal, suggesting that cross-asset interactions can be directional and temporally shifted rather than perfectly synchronized.

\begin{figure}[t]
\centering
\includegraphics[width=1.0\columnwidth]{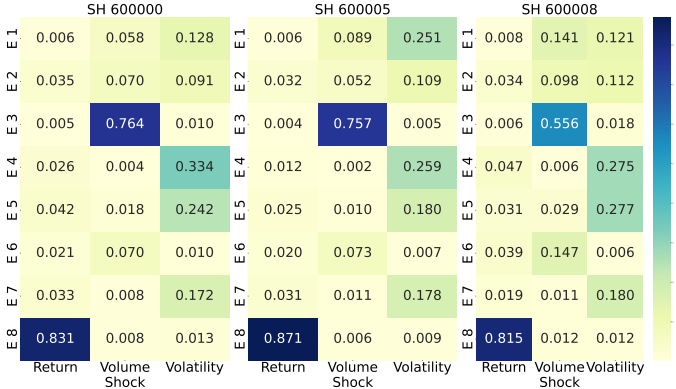} 
\caption{MoE routing weights across tasks on March 20, 2020. The x-axis shows the three tasks, and the y-axis shows the routing weights.}
\label{fig:moe_heatmap}
\end{figure}

Finally, Figure~\ref{fig:moe_heatmap} visualizes the gating weights of three representative stocks (\texttt{SH600000}, \texttt{SH600005}, and \texttt{SH600008}) across the Return, Volume Shock, and Volatility tasks. Because experts are shared across objectives, the distinct routing patterns across tasks illustrate how LDL allocates expert capacity differently for return, liquidity, and risk signals.

\subsection{Ablation Study (RQ2)}
To validate the effectiveness of each proposed component, we conduct systematic ablation studies by progressively removing key modules. Table~\ref{tab:ablation} summarizes the results on CSI300 data.

\begin{itemize}
    \item \textbf{w/o Auxiliary Tasks:} This variant removes the volume and volatility objectives and trains only on return prediction. Performance drops across all metrics, by 1.39\,\% to 7.80\,\%, showing that the auxiliary targets provide useful supervision beyond simple regularization.

    \item \textbf{w/o LDL:} This variant removes the LDL transfer module and feeds the MRE output directly to the prediction heads. The resulting gap to \model~shows that simply adding auxiliary heads is not sufficient; gated cross-task transfer is a key part of the improvement.

    \item \textbf{w/o Auxiliary Tasks \& LDL:} This variant removes both the auxiliary tasks and LDL. The broader degradation confirms that joint supervision and cross-task transfer are complementary.

    \item \textbf{w/o Cross-Stock Modeling:} This variant disables the cross-stock attention layer and therefore cannot use contemporaneous peer information. It causes the largest IC drop, from 0.0575 to 0.0493, supporting the importance of explicit cross-sectional interaction.

    \item \textbf{w/o Auxiliary Tasks \& LDL \& Cross-Stock Modeling:} This temporal-only variant performs worst overall, indicating that the gains do not come from model scale alone but from the combination of cross-stock modeling and liquidity-aware multi-task learning.
    
\end{itemize}

To verify that the auxiliary objectives provide informative supervision rather than acting as mere regularizers, we also evaluate the auxiliary heads directly on the held-out split. The volume-shock task achieves IC/ICIR of 0.3435/2.95, while the volatility task achieves IC/ICIR of 0.5157/4.36. These results indicate that LDL learns auxiliary liquidity and risk signals that are predictive in their own right, which helps explain the gains from cross-task transfer to the return objective and, ultimately, to APO.

\subsection{Backtest Evaluation (RQ3)}

To evaluate deployment value, we conduct backtests on CSI300 over 2017--2020. The strategy uses return forecasts for stock selection and volume/volatility forecasts for liquidity-aware position sizing.

\subsubsection{\textbf{Experiment Setting}}
We form daily long-short portfolios by sorting stocks into return-based deciles, going long the top decile and short the bottom decile, and assigning weights with APO. Table~\ref{tab:backtest_comparison} compares APO against equal-weighted, inverse-volatility risk-parity, and rank-based linear-weighting baselines; the exact baseline formulas are given in Appendix~\ref{sec:apo_baselines}. Each side is allocated CNY $1\times 10^9$, transaction cost is 10 basis points per trade, and each stock is capped at 2\% of daily turnover.

\subsubsection{\textbf{Experiment Performance}}
\begin{figure}[t]
\centering
\includegraphics[width=1.0\columnwidth]{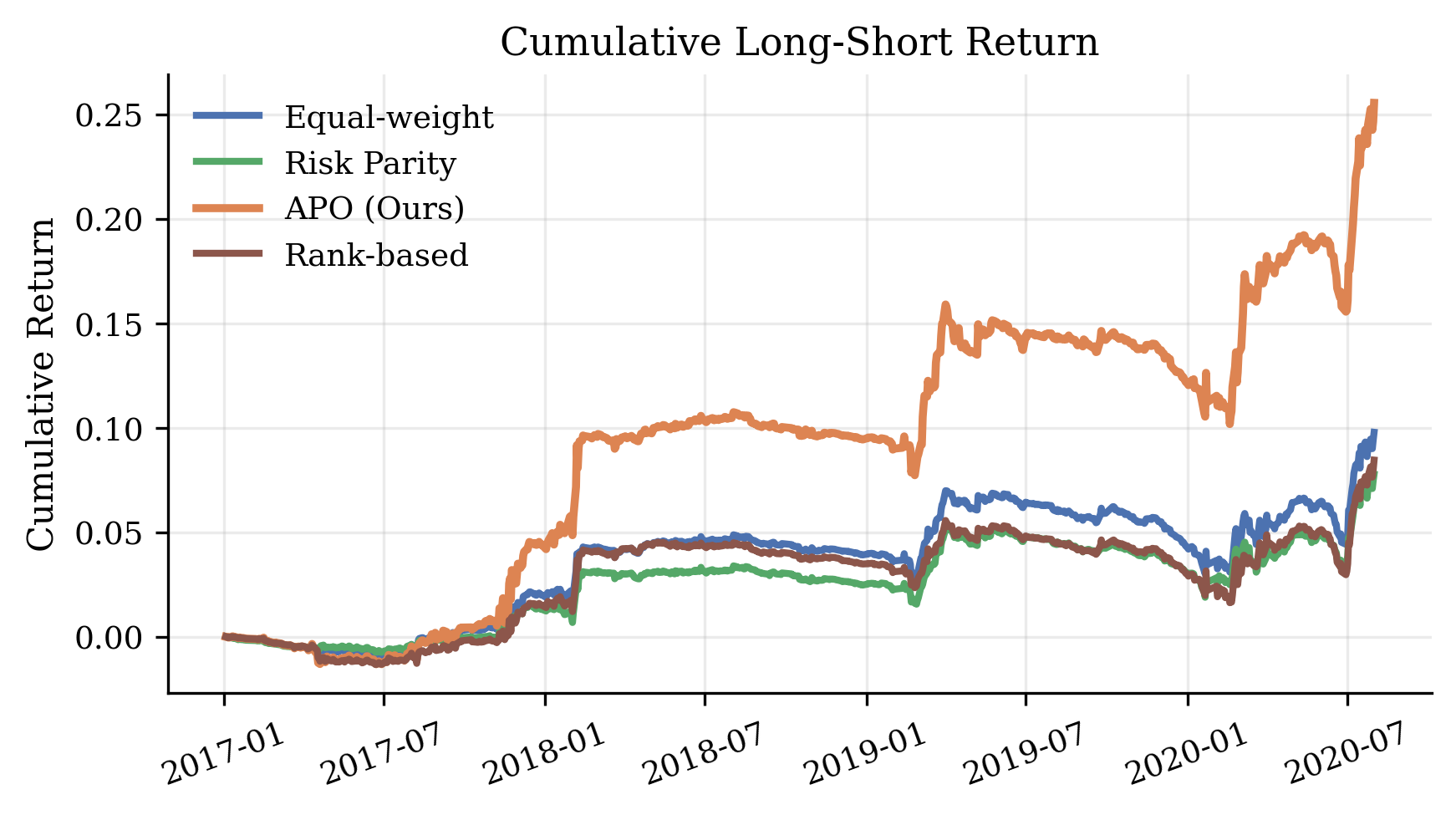} 
\caption{Cumulative long-short returns of APO and three portfolio-construction baselines.}
\label{figG}
\end{figure}

Table~\ref{tab:backtest_comparison} shows that APO (with $r=0.1$) delivers the strongest profitability under the tested trading constraints. Relative to equal weighting, annualized return rises from 3.99\% to 10.01\%, OSR from 32.13\% to 38.00\%, and Sharpe ratio from 1.22 to 1.86; Sortino and Calmar likewise improve to 2.00 and 2.03. APO also exceeds the rank-based baseline, lifting annualized return from 3.38\% to 10.01\%, OSR from 30.05\% to 38.00\%, and Sharpe ratio from 1.18 to 1.86. These gains come with higher drawdown and turnover, indicating a deliberate return--risk trade-off rather than mechanically lower-risk allocations. The underlying model-induced long-short spread also remains significant after standard risk adjustment: over the same 2017--2020 window, a Fama-French five-factor regression still yields a daily alpha of 0.35\% ($p<0.01$). This suggests that APO improves how the signal is executed, rather than merely amplifying conventional style exposures. Figure~\ref{figG} shows that APO remains above the rank-based portfolio for most of the sample, with a particularly visible separation during the February--March 2020 stress episode.

% \begin{figure}[t]
% \centering
% \includesvg[width=0.9\columnwidth]{Pictures/Combined_Charts.svg} 
% \caption{Comparative analysis of three key indicators.}
% \label{figC}
% \end{figure}

% \subsubsection{Economic Interpretation}

\begin{table}[t]
\centering
\renewcommand\arraystretch{1.0}
\footnotesize
\caption{Performance comparison: APO vs. traditional portfolio construction baselines}
\label{tab:backtest_comparison}
\setlength{\tabcolsep}{3pt}
\resizebox{\columnwidth}{!}{%
\begin{tabular}{l|cccc}
\toprule
\textbf{Metric} 
& \textbf{Equal-weighted} 
& \textbf{Risk Parity} 
& \textbf{Rank-based}
& \textbf{APO (Ours, $r=0.1$)} \\
\midrule
OSR$\uparrow$        & 32.13\%  & 29.27\%  & 30.05\% & 38.00\% \\
AR$\uparrow$         & 3.99\%   & 3.19\%   & 3.38\%  & 10.01\% \\
Downside Vol$\downarrow$ & 0.031 & 0.027 & 0.0278 & 0.050 \\
MD$\downarrow$       & 3.73\%   & 3.15\%   & 3.28\%  & 4.94\% \\
Turnover$\downarrow$ & 20.73\% & 19.13\% & 19.42\% & 24.98\% \\
ShR$\uparrow$        & 1.22     & 1.16     & 1.18     & 1.86 \\
P/L$\uparrow$        & 1.59     & 1.69     & 1.67     & 1.73 \\
SoR$\uparrow$        & 1.29     & 1.20     & 1.22     & 2.00 \\
CR$\uparrow$         & 1.07     & 1.01     & 1.03     & 2.03 \\
\bottomrule
\end{tabular}
}
\end{table}

\subsubsection{\textbf{Parameter Analysis}}

We evaluate the sensitivity of APO to $r \in [0.1, 0.9]$. Table~\ref{tab:sensitivity1} reports the resulting return, risk, and execution metrics.

\begin{table*}[t]
\centering
\renewcommand\arraystretch{1.0}
\footnotesize
\caption{Sensitivity analysis of weighting parameter $r$.}
\label{tab:sensitivity1}
\setlength{\tabcolsep}{4pt}
\resizebox{\textwidth}{!}{%
\begin{tabular}{c|ccccccccc}
\toprule
$r$ 
& \textbf{OSR$\uparrow$} 
& \textbf{AR$\uparrow$} 
& \textbf{ShR$\uparrow$} 
& \textbf{MD$\downarrow$} 
& \textbf{P/L Ratio$\uparrow$} 
& \textbf{SoR$\uparrow$} 
& \textbf{CR$\uparrow$} 
& \textbf{Downside Vol$\downarrow$} 
& \textbf{Turnover Ratio$\downarrow$} \\
\midrule
0.1 & \textbf{0.3800} & \textbf{0.1001} & \textbf{1.8607} & 0.0494 & \textbf{1.7309} & \textbf{2.0030} & \textbf{2.0260} & 0.0500 & 0.2498 \\
0.2 & 0.3774 & 0.0933 & 1.8316 & 0.0470 & 1.7100 & 1.9895 & 1.9864 & 0.0469 & 0.2479 \\
0.3 & 0.3737 & 0.0848 & 1.7903 & 0.0438 & 1.6672 & 1.9515 & 1.9367 & 0.0435 & 0.2453 \\
0.4 & 0.3684 & 0.0765 & 1.7649 & 0.0408 & 1.6510 & 1.9410 & 1.8738 & 0.0394 & 0.2416 \\
0.5 & 0.3611 & 0.0685 & 1.7494 & 0.0385 & 1.6397 & 1.9501 & 1.7777 & 0.0351 & 0.2366 \\
0.6 & 0.3509 & 0.0590 & 1.6854 & 0.0373 & 1.6560 & 1.8827 & 1.5807 & 0.0314 & 0.2299 \\
0.7 & 0.3367 & 0.0480 & 1.5488 & 0.0355 & 1.6279 & 1.7004 & 1.3542 & 0.0283 & 0.2207 \\
0.8 & 0.3171 & 0.0372 & 1.3691 & 0.0320 & 1.6192 & 1.4652 & 1.1621 & 0.0254 & 0.2086 \\
0.9 & 0.2913 & 0.0262 & 1.1035 & \textbf{0.0282} & 1.6274 & 1.1405 & 0.9297 & \textbf{0.0230} & \textbf{0.1937} \\
\bottomrule
\end{tabular}
}
\end{table*}

The best return and risk-adjusted performance occur at $r=0.1$. As $r$ increases, APO places more weight on volatility control, reducing turnover and drawdown but also weakening profitability, which indicates that a mild liquidity preference is most effective in this market.

\section{Conclusions and Future Work}
We presented LiMT, a hierarchical multi-task framework for stock forecasting and liquidity-aware portfolio construction. LiMT combines alternating cross-stock/temporal attention, cross-task expert routing, and a lightweight APO rule. Across CSI300 and CSI500, it achieves the best results among the compared baselines; in realistic CSI300 backtests, APO improves annualized return from 3.99\% to 10.01\% and Sharpe ratio from 1.22 to 1.86 over equal weighting. These results suggest that jointly modeling return, volume, and volatility can improve both predictive quality and executable portfolio construction.

Future work includes extending the framework to intraday microstructure data, incorporating richer alternative data sources, and modeling cross-market spillovers.

\appendices
\section{Evaluation and Implementation} \label{sec:EI}
\textbf{Implementation.} We implement \model~in PyTorch on top of Qlib~\cite{yang2020qlib} and use Qlib implementations for all baselines. For baseline tuning, the number of layers and hidden size are searched over $\{1,2,3\}$ and $\{128,256,512\}$, and the learning rate over $\{10^{-i}\}_{i\in\{3,4,5,6\}}$. For \model, we tune the hidden size and learning rate in the same ranges, and use $D=256$, $lr=10^{-5}$, $N_1=4$, and $N_2=2$ across datasets. The task weights $\lambda_r$, $\lambda_v$, and $\lambda_\sigma$ are selected from $\{0.63,0.65,0.68,0.7\}$, $\{0.3,0.33,0.35\}$, and $\{0.03,0.05,0.08\}$ according to validation IC.

Each model is trained for at most 100 epochs with early stopping on a server with an Intel Xeon Platinum 8358 CPU, 2TB memory, and one RTX3090 GPU (24GB memory). Every experiment is repeated three times with different random initializations, and the average performance is reported. For APO backtesting, we allocate CNY $1\times 10^{9}$ to each of the long and short sides, set transaction cost to 10 basis points per trade, and cap the traded value of each stock at 2\% of its daily turnover.

\subsection{Backtest Portfolio Baselines}
\label{sec:apo_baselines}
To make the portfolio-construction comparison reproducible, we define the three non-APO weighting rules used in Figure~\ref{figG}. Let $\mathcal{G}_t$ denote the candidate stock set on trading day $t$ (corresponding to the long or short decile), let $C_{i,t}$ be the close price of stock $i$, and let $s_{i,t}$ be its predicted return score.

\textbf{Equal-weighted:}
\begin{equation}
w_{i,t}^{\mathrm{EW}}=\frac{1}{|\mathcal{G}_t|},\quad i\in \mathcal{G}_t .
\end{equation}

\textbf{Risk parity (inverse-volatility):} we first compute one-day returns and 20-day historical volatility,
\begin{equation}
r_{i,t}^{(1d)}=\frac{C_{i,t}}{C_{i,t-1}}-1,\qquad
\hat{\sigma}_{i,t}=\mathrm{Std}\!\left(r_{i,t-k}^{(1d)}\right)_{k=0}^{19},
\end{equation}
then allocate weights by normalized inverse volatility:
\begin{equation}
w_{i,t}^{\mathrm{RP}}=\frac{\hat{\sigma}_{i,t}^{-1}}{\sum_{j\in\mathcal{G}_t}\hat{\sigma}_{j,t}^{-1}},\quad i\in \mathcal{G}_t .
\end{equation}

\textbf{Rank-based (linear normalization):} let $\rho_{i,t}=\mathrm{Rank}_{\mathcal{G}_t}(s_{i,t})$. We reverse the ranks for the short group and keep them unchanged for the long group:
\begin{equation}
\tilde{\rho}_{i,t}=
\begin{cases}
|\mathcal{G}_t|-\rho_{i,t}+1, & \mathcal{G}_t=\mathcal{G}_t^{\mathrm{short}},\\
\rho_{i,t}, & \mathcal{G}_t=\mathcal{G}_t^{\mathrm{long}},
\end{cases}
\end{equation}
followed by linear normalization:
\begin{equation}
w_{i,t}^{\mathrm{RK}}=
\frac{\tilde{\rho}_{i,t}-\min_{j\in\mathcal{G}_t}\tilde{\rho}_{j,t}}
{\sum_{j\in\mathcal{G}_t}\left(\tilde{\rho}_{j,t}-\min_{k\in\mathcal{G}_t}\tilde{\rho}_{k,t}\right)},
\quad i\in \mathcal{G}_t .
\end{equation}

\section{Metric Definitions}
For predictive evaluation, IC measures the daily correlation between predicted scores and realized returns, and ICIR normalizes IC by its temporal standard deviation. For benchmark-relative portfolio evaluation, AR$_{\text{excess}}$ and IR$_{\text{excess}}$ measure excess profitability and its stability against the corresponding CSI benchmark.

For APO backtests, we additionally report annualized return, Sharpe ratio, Sortino ratio, Calmar ratio, maximum drawdown, downside volatility, turnover, order success rate, and profit-to-loss ratio to summarize profitability, downside risk, and execution quality.

\section{Computational Complexity Analysis}
Let $S$ denote the number of stocks, $T$ the lookback length, $d$ the hidden dimension, $d_k$ the attention-head dimension, and $n$ the number of experts. Cross-stock attention costs $O(TS^2d_k)$, temporal attention costs $O(ST^2d_k)$, and expert routing/aggregation costs $O(Sdn)$. Hence the total time complexity is dominated by
\begin{equation}
O\!\left(S\max(T^2,S)d_k + Sdn\right),
\end{equation}
with space complexity
\begin{equation}
O\!\left(TS^2 + ST^2 + STd\right).
\end{equation}
This is substantially smaller than direct pairwise cross-time interaction, which would scale as $O(S^2T^2d_k)$.

\bibliographystyle{IEEEtran}
\bibliography{Reference}

@String{Computing = "Computing" }

@inproceedings{yoo2021accurate,
  title={Accurate multivariate stock movement prediction via data-axis transformer with multi-level contexts},
  author={Yoo, Jaemin and Soun, Yejun and Park, Yong-chan and Kang, U},
  booktitle={Proceedings of the 27th ACM SIGKDD Conference on Knowledge Discovery \& Data Mining},
  pages={2037--2045},
  year={2021}
}

@article{parkStockMarketForecasting2022,
  title={Stock market forecasting using a multi-task approach integrating long short-term memory and the random forest framework},
  author={Park, Hyun Jun and Kim, Youngjun and Kim, Ha Young},
  journal={Applied Soft Computing},
  volume={114},
  pages={108106},
  year={2022},
  publisher={Elsevier}
}

@inproceedings{maMultipleStockTime2020a,
  title = {Multiple {{Stock Time Series Jointly Forecasting}} with {{Multi-Task Learning}}},
  booktitle = {2020 {{International Joint Conference}} on {{Neural Networks}} ({{IJCNN}})},
  author = {Ma, Tao and Tan, Ying},
  year = {2020},
  month = jul,
  pages = {1--8},
  issn = {2161-4407},
  doi = {10.1109/IJCNN48605.2020.9207543},
  urldate = {2025-04-09}
}

@misc{goyenkoTradingVolumeAlpha2024,
  type = {{{SSRN Scholarly Paper}}},
  title = {Trading {{Volume Alpha}}},
  author = {Goyenko, Ruslan and Kelly, Bryan T. and Moskowitz, Tobias J. and Su, Yinan and Zhang, Chao},
  year = {2024},
  month = may,
  number = {4802345},
  eprint = {4802345},
  publisher = {Social Science Research Network},
  address = {Rochester, NY},
  doi = {10.2139/ssrn.4802345},
  urldate = {2025-04-08},
  archiveprefix = {Social Science Research Network},
  langid = {english}
}

@inproceedings{liMASTERMarketGuidedStock2023,
  title={Master: Market-guided stock transformer for stock price forecasting},
  author={Li, Tong and Liu, Zhaoyang and Shen, Yanyan and Wang, Xue and Chen, Haokun and Huang, Sen},
  booktitle={Proceedings of the AAAI conference on artificial intelligence},
  volume={38},
  number={1},
  pages={162--170},
  year={2024}
}

@article{tcn,
  title={An empirical evaluation of generic convolutional and recurrent networks for sequence modeling},
  author={Bai, Shaojie and Kolter, J Zico and Koltun, Vladlen},
  journal={arXiv},
  year={2018}
}

@inproceedings{tabnet,
  title={Tabnet: Attentive interpretable tabular learning},
  author={Arik, Sercan {\"O} and Pfister, Tomas},
  booktitle={AAAI},
  volume={35},
  number={8},
  pages={6679--6687},
  year={2021}
}

@inproceedings{gru,
  author = {Kyunghyun Cho and Bart van Merri{\"e}nboer and Dzmitry Bahdanau and Yoshua Bengio},
  title = {On the Properties of Neural Machine Translation: Encoder-Decoder Approaches},
  booktitle = {SSST},
  year = {2014},
  pages = {103--111}
}

@article{lstm,
  title={Long Short-term Memory},
  author={Hochreiter, S},
  journal={Neural Computation MIT-Press},
  year={1997}
}

@article{yadati2025localformer,
  title={Localformer: Mitigating over-globalising in transformers on graphs with localised training},
  author={Yadati, Naganand},
  journal={Transactions on Machine Learning Research},
  year={2025}
}

@inproceedings{sfm,
    author = {Zhang, Liheng and Aggarwal, Charu and Qi, Guo-Jun},
    title = {Stock Price Prediction via Discovering Multi-Frequency Trading Patterns},
    year = {2017},
    booktitle = {SIGKDD},
    pages = {2141–2149},
}

@article{TFT,
  author = {Bryan Lim and Sercan O. Arik and Nicolas Loeff and Tomas Pfister},
  title = {Temporal Fusion Transformers for Interpretable Multi-horizon Time Series Forecasting},
  journal = {IJF},
  year = {2021},
  volume = {37},
  number = {4},
  pages = {1748--1764}
}

@inproceedings{qin_dual-stage_2017,
    location = {Melbourne, Australia},
    title = {A Dual-Stage Attention-Based Recurrent Neural Network for Time Series Prediction},
    booktitle = {IJCAI},
    pages = {2627–2633},
    year = {2017},
    numpages = {7},
    author = {Qin, Yao and Song, Dongjin and Chen, Haifeng and Cheng, Wei and Jiang, Guofei and Cottrell, Garrison W.}
}

@article{feng2019temporal,
  title={Temporal relational ranking for stock prediction},
  author={Feng, Fuli and He, Xiangnan and Wang, Xiang and Luo, Cheng and Liu, Yiqun and Chua, Tat-Seng},
  journal={ACM Transactions on Information Systems (TOIS)},
  volume={37},
  number={2},
  pages={1--30},
  year={2019},
  publisher={ACM New York, NY, USA}
}

@inproceedings{chen2016xgboost,
  title={Xgboost: A scalable tree boosting system},
  author={Chen, Tianqi and Guestrin, Carlos},
  booktitle={Proceedings of the 22nd acm sigkdd international conference on knowledge discovery and data mining},
  pages={785--794},
  year={2016}
}

@article{ke2017lightgbm,
  title={Lightgbm: A highly efficient gradient boosting decision tree},
  author={Ke, Guolin and Meng, Qi and Finley, Thomas and Wang, Taifeng and Chen, Wei and Ma, Weidong and Ye, Qiwei and Liu, Tie-Yan},
  journal={Advances in neural information processing systems},
  volume={30},
  year={2017}
}

@inproceedings{zhang2020doubleensemble,
  title={Doubleensemble: A new ensemble method based on sample reweighting and feature selection for financial data analysis},
  author={Zhang, Chuheng and Li, Yuanqi and Chen, Xi and Jin, Yifei and Tang, Pingzhong and Li, Jian},
  booktitle={2020 IEEE international conference on data mining (ICDM)},
  pages={781--790},
  year={2020},
  organization={IEEE}
}

@article{engleMultipleIndicatorsModel2006,
  title = {A Multiple Indicators Model for Volatility Using Intra-Daily Data},
  author = {Engle, Robert F. and Gallo, Giampiero M.},
  year = {2006},
  month = mar,
  journal = {Journal of Econometrics},
  volume = {131},
  number = {1-2},
  pages = {3--27},
  issn = {03044076},
  doi = {10.1016/j.jeconom.2005.01.018},
  urldate = {2025-07-30},
  copyright = {https://www.elsevier.com/tdm/userlicense/1.0/},
  langid = {english}
}

@article{zhuPredictiveRegressionsMacroeconomic2014,
  title = {Predictive Regressions for Macroeconomic Data},
  author = {Zhu, Fukang and Cai, Zongwu and Peng, Liang},
  year = {2014},
  month = mar,
  journal = {The Annals of Applied Statistics},
  volume = {8},
  number = {1},
  pages = {577--594},
  publisher = {Institute of Mathematical Statistics},
  issn = {1932-6157, 1941-7330},
  doi = {10.1214/13-AOAS708},
  urldate = {2025-08-01}
}

@misc{khanMacroeconomicFactorsStock2023,
  title = {Macroeconomic Factors and {{Stock}} Exchange Return: {{A Statistical Analysis}}},
  shorttitle = {Macroeconomic Factors and {{Stock}} Exchange Return},
  author = {Khan, Md Fazlul Huq and Billah, Md Masum},
  year = {2023},
  month = may,
  number = {arXiv:2305.02229},
  eprint = {2305.02229},
  primaryclass = {econ},
  publisher = {arXiv},
  doi = {10.48550/arXiv.2305.02229},
  urldate = {2025-08-01},
  archiveprefix = {arXiv}
}

@article{szucsForecastingIntradayVolume2017,
  title = {Forecasting Intraday Volume: {{Comparison}} of Two Early Models},
  shorttitle = {Forecasting Intraday Volume},
  author = {Sz{\H u}cs, Bal{\'a}zs {\'A}rp{\'a}d},
  year = {2017},
  month = may,
  journal = {Finance Research Letters},
  volume = {21},
  pages = {249--258},
  issn = {1544-6123},
  doi = {10.1016/j.frl.2016.11.018},
  urldate = {2025-07-30}
}

@book{zhou2012ensemble,
    title={Ensemble methods: foundations and algorithms},
    author={Zhou, Zhi-Hua},
    year={2012},
    publisher = {Chapman \& Hall/CRC},
}

@inproceedings{selvin2017stock,
  title={Stock price prediction using LSTM, RNN and CNN-sliding window model},
  author={Selvin, Sreelekshmy and Vinayakumar, Ravi and Gopalakrishnan, EA and Menon, Vijay Krishna and Soman, KP},
  booktitle={2017 international conference on advances in computing, communications and informatics (icacci)},
  pages={1643--1647},
  year={2017},
  organization={IEEE}
}

@inproceedings{du2021adarnn,
  title={Adarnn: Adaptive learning and forecasting of time series},
  author={Du, Yuntao and Wang, Jindong and Feng, Wenjie and Pan, Sinno and Qin, Tao and Xu, Renjun and Wang, Chongjun},
  booktitle={Proceedings of the 30th ACM international conference on information \& knowledge management},
  pages={402--411},
  year={2021}
}

@inproceedings{zhang2017stock,
  title={Stock price prediction via discovering multi-frequency trading patterns},
  author={Zhang, Liheng and Aggarwal, Charu and Qi, Guo-Jun},
  booktitle={SIGKDD},
  pages={2141--2149},
  year={2017}
}

@inproceedings{nettlesForecastingIntradayVolume2015,
  title = {Forecasting Intraday Volume Distributions},
  booktitle = {2015 {{Systems}} and {{Information Engineering Design Symposium}}},
  author = {Nettles, James and Brayer, Nicholas and Jenner, Charles and Ngo, Alexander and Putnam, Charlie and Shank, Trevor and Todd, Andrew and Beling, Peter},
  year = {2015},
  month = apr,
  pages = {97--102},
  doi = {10.1109/SIEDS.2015.7117019},
  urldate = {2025-07-30}
}

@misc{zhangVolatilityForecastingMachine2022,
  type = {{{SSRN Scholarly Paper}}},
  title = {Volatility {{Forecasting}} with {{Machine Learning}} and {{Intraday Commonality}}},
  author = {Zhang, Chao and Zhang, Yihuang and Cucuringu, Mihai and Qian, Zhongmin},
  year = {2022},
  month = may,
  number = {4022147},
  eprint = {4022147},
  publisher = {Social Science Research Network},
  address = {Rochester, NY},
  doi = {10.2139/ssrn.4022147},
  urldate = {2025-07-30},
  archiveprefix = {Social Science Research Network},
  langid = {english}
}

@inproceedings{wang2021hierarchical,
  title={Hierarchical Adaptive Temporal-Relational Modeling for Stock Trend Prediction.},
  author={Wang, Heyuan and Li, Shun and Wang, Tengjiao and Zheng, Jiayi},
  booktitle={IJCAI},
  pages={3691--3698},
  year={2021}
}

@inproceedings{wang2022adaptive,
  title={Adaptive long-short pattern transformer for stock investment selection},
  author={Wang, Heyuan and Wang, Tengjiao and Li, Shun and Zheng, Jiayi and Guan, Shijie and Chen, Wei},
  booktitle={Proceedings of the Thirty-First International Joint Conference on Artificial Intelligence},
  pages={3970--3977},
  year={2022}
}

@inproceedings{xiang2022temporal,
  title={Temporal and Heterogeneous Graph Neural Network for Financial Time Series Prediction},
  author={Xiang, Sheng and Cheng, Dawei and Shang, Chencheng and Zhang, Ying and Liang, Yuqi},
  booktitle={Proceedings of the 31st ACM International Conference on Information \& Knowledge Management},
  pages={3584--3593},
  year={2022}
}

@article{yang2020qlib,
  title={Qlib: An ai-oriented quantitative investment platform},
  author={Yang, Xiao and Liu, Weiqing and Zhou, Dong and Bian, Jiang and Liu, Tie-Yan},
  journal={arXiv preprint arXiv:2009.11189},
  year={2020}
}

@article{vaswani2017attention,
  title={Attention is all you need},
  author={Vaswani, Ashish and Shazeer, Noam and Parmar, Niki and Uszkoreit, Jakob and Jones, Llion and Gomez, Aidan N and Kaiser, {\L}ukasz and Polosukhin, Illia},
  journal={Advances in neural information processing systems},
  volume={30},
  year={2017}
}

@article{catboost,
  author = {Liudmila Prokhorenkova and Gleb Gusev and Aleksandr Vorobev and Anna Veronika Dorogush and Andrey Gulin},
  title = {CatBoost: Unbiased Boosting with Categorical Features},
  journal = {NeurIPS},
  volume={31},
  year={2018}
}

@article{hoseinzade2019cnnpred,
  title={CNNpred: CNN-based stock market prediction using a diverse set of variables},
  author={Hoseinzade, Ehsan and Haratizadeh, Saman},
  journal={ESA},
  volume={129},
  pages={273--285},
  year={2019}
}

@article{quinlan1986induction,
  title={Induction of decision trees},
  author={Quinlan, J. Ross},
  journal={ML},
  pages={81--106},
  year={1986},
}

@book{quinlan2014c4,
  title={C4. 5: programs for machine learning},
  author={Quinlan, J Ross},
  year={2014},
  publisher={Elsevier}
}

@inproceedings{linCSPOCrossMarketSynergistic2025,
  title={Cspo: Cross-market synergistic stock price movement forecasting with pseudo-volatility optimization},
  author={Lin, Sida and Chen, Yankai and Qi, Yiyan and Ma, Chenhao and Cao, Bokai and Zhang, Yifei and Liu, Xue and Guo, Jian},
  booktitle={Companion Proceedings of the ACM on Web Conference 2025},
  pages={354--363},
  year={2025}
}

\end{document}